# **Early maturation** processes in coal.

## Part 1: Pyrolysis mass balances and structural evolution of coalified wood from the Morwell Brown Coal seam

Elodie Salmon [a, c], Françoise Behar [a], François Lorant [a], Patrick G. Hatcher [b], Paul-Marie Marquaire [c].

*a Institut Français du Pétrole, BP 311, 92506 Rueil-Malmaison cedex, France*

*b Department of Chemistry and Biochemistry, Old Dominion University, Norfolk, VA, USA 23529*

*c Département de Chimie Physique des Réactions, Institut Polytechnique de Lorraine, 1 rue Grandville, B.P.20451 F, 54001 Nancy cedex, France.*

Corresponding author:

François Lorant, IFP (Institut Français du Pétrole), Département de Géochimie, 1 et 4 avenue du bois Préau, 92852 Rueil-Malmaison, France, tel : 33 14 752 6910 fax : 33 14 752 7019, email : francois.lorant@ifp.fr

## Abstract

In this work, we develop a theoretical approach to evaluate maturation process of kerogen-like material, involving molecular dynamic reactive modeling with a reactive force field to simulate the thermal stress. The Morwell coal has been selected to study the thermal evolution of terrestrial organic matter. To achieve this, a structural model is first constructed based on models from the literature and analytical characterization of our samples by modern 1-and 2-D NMR, FTIR, and elemental analysis. Then, artificial maturation of the Morwell coal is performed at low conversions in order to obtain, quantitative and qualitative, detailed evidences of structural evolution of the kerogen upon maturation. **The observed chemical changes are a defunctionalization of the carboxyl, carbonyl and methoxy functional groups coupling with an increase of cross linking in the residual mature kerogen. Gaseous and liquids hydrocarbons, essentially $CH_4$, $C_4H_8$ and $C_{14+}$ liquid hydrocarbons, are generated in low amount, merely by cleavage of the lignin side chain.**

## 1. **Introduction:**

Thermal stress in the Earth's subsurface is one of the most important forces driving hydrocarbon generation from kerogen in shales and coal (Philippi, 1965; Louis **and** Tissot, 1967; Albrecht and Ourisson, 1969). For terrestrial material that typically forms coal, the main **hydrocarbon** is methane gas. However, in some cases, coal is considered to be a source of paraffinic-rich hydrocarbons of **molecular weight higher** than methane or volatile hydrocarbon gases (Mukhopadhyay et al., 1991; Fowler et al., 1991, Nelson et al., 1998). Of particular interest in this current study is the origin of methane and other gaseous hydrocarbons, especially from precursor chemical structures like lignin. Previous studies have shown that lignin is an important precursor for coal structures (Hatcher, 1989) and that methane is generated in abundance as the main hydrocarbon from such a structural component of wood during maturation (Behar and Hatcher, 1995). Moreover, here are many kerogens whose organic matter is partly sourced from terrestrially-derived organic matter which contains as its main constituent lignin. Knowing the hydrocarbon-generating potential and mechanism from the lignaceous components of such kerogens is **of paramount interest** to assessing the relative importance of terrestrial organic matter in the overall hydrocarbon potential of these kerogens **(Behar and Hatcher, 1995)**.

The work reported in the present paper is part of a study that aims at establishing, from a theoretical point of view, the primary cracking mechanism for **insoluble sedimentary organic matter** derived exclusively from lignin. This paper addresses the structural study of specific **insoluble sedimentary organic matter** materials and the construction of the corresponding molecular models. We also describe the thermal decomposition, at low conversions, of a sample in which we deduce the initial chemical reactions for conversion. For this purpose, a lignitic wood from the Morwell coal (Victorian brown coal, Australia) was selected as our precursor material, recognizing that it has already undergone some maturation to achieve a coal rank of lignite **(Hatcher, 1988)**. This wood sample has previously been shown by flash pyrolysis/gas chromatography/mass spectrometry to contain mainly lignin-derived structures (Behar and Hatcher, 1995). In a previous study, the chemical structural composition and a proposed structural model for gymnospermous wood was presented (Hatcher, 1989). However, a model for angiospermous lignite has yet to be proposed.

Our approach is to first compare our chemical analyses to chemical models previously proposed in the literature for similar samples. Based on both literature models and our own analyses, updated chemical structures are proposed for a material derived from angiospermous wood. Closed pyrolysis of the wood in gold tubes followed by a detailed quantitative analysis of the products, including the matured lignite residue, allows for mass balances of products and determination of the overall transformation processes during simulated maturation. Thermal decomposition products are identified using numerous techniques such as elemental analyses, gas chromatography (GC) coupled to flame ionization detection (FID) or thermal conductivity detection (TCD), infrared (IR) spectroscopy, and solid and liquid state $^{13}C$ NMR.

## 2. Sample: Lignite from the Morwell coal

**The Morwell lignite was collected from the Morwell Open Cut, Latrobe Valley, Victoria, Australia and is composed essentially of angiospermous wood transformed to the rank of lignite B (Hatcher, 1988; Behar and Hatcher, 1995).** It is part of an entire fossil tree buried at the peat stage in the Early Miocene and coalified within the deposit. During early diagenesis, biodegradation and mild chemical processes led to the decomposition of the cellulose and the hemi-cellulose and transformed the lignin structure into lignite (Spackman and Barghoorn, 1966; Philp et al., 1982; Stout et al., 1988; Hatcher et al., 1989, Hatcher and Clifford, 1997). The sample was freeze-dried and ground to a fine

powder with a mortar and pestle and stored under nitrogen gas. **This sample is ideal for this study because it is naturally organic rich, no chemical treatment is needed to extract the organic matter and it is thermally immature.**

## 3. Experimental

The experimental procedures used in this study have been described in detail by Salmon et al. (**accepted manuscript**). In this previous work, similar experiments were performed on an aliphatic biopolymer, algaenan from *Botryococcus braunii* race L. The structural evolution of the Morwell lignite was performed by characterizing the initial sample and the recovered residues from simulated maturations at various temperatures and times. Elemental analysis (combustion/pyrolysis in a Carlo-Erba system) is used to quantify the proportions of C, H and O. Chemical functional groups are measured by Attenuated Total Reflectance Fourier Transform Infrared (ATR – FTIR) spectroscopy on a Bruker Tenser 27 spectrometer.

Detailed structural characterizations are obtained by direct polarisation and magic angle spinning (DPMAS) $^{13}C$ NMR and High Resolution Magic Angle Spinning (HRMAS) NMR techniques on a Bruker AVANCE II Ultra Shield™ 400 MHz spectrometer. Solid-state $^{13}C$-NMR spectra were obtained using the basic direct polarization pulse program as described previously (Dria et al., 2002). Approximately 80 mg of sample was inserted into an NMR rotor and spun at the magic angle ($54.7^{o}$) with a frequency of 15 kHz. A 45 degree pulse angle and a 10 s recycle time were used for each of 10,000 accumulations. Exactly 1024 data points were collected on the free induction decay and an exponential line broadening of 100 Hz was applied prior to Fourier transformation. The spectra were integrated by dropping vertical lines to the baseline between chemical shift regions characteristic of the various types of functional groups.

HRMAS spectroscopy was performed with the same NMR spectrometer as described above using a CHN z-axis gradient HRMAS probe. Approximately 20 mg of sample was swelled in DMSO-$d_6$ (Aldrich, 99.9 atom % D) as it was packed into a 4 mm diameter zirconia MAS rotor spun at the magic angle at 9 kHz. A relaxation delay of 1 s was used for each experiment. A $^{1}H$-$^{13}C$ heteronuclear single quantum coherence (HSQC) spectrum was acquired using echo-antiecho gradient selection. In the $^{1}H$ dimension (F2), 344 scans were acquired, each collected with 1024 data points for a spectral width of 4,006 Hz (10.01 ppm). In the $^{13}C$ dimension (F1), 128 data points were collected for a spectral width of 166 ppm. Line broadening was used in both dimensions, 1 Hz in the F1 and 0.3 Hz in the F2 dimension. The FIDs were processed in both dimensions using a squared sine multiplication (QSINE) window function.

A total correlation spectroscopy (TOCSY) spectrum was acquired with a phase sensitive pulse program that used States-TPPI and the MLEV-17 multiple pulse spin lock sequence. A mixing time of 60 ms was used. A spectral width of 6000 Hz (15 ppm) was obtained in both dimensions. In the F2 dimension, 128 scans were acquired, with 2048 data points. In the F1 dimension, 256 data points were collected and zero-filled to 512. The FIDs were processed in the F2 dimensions using a -3 Hz Gaussian line broadening and in the F1 dimension using a line broadening of 1.0 and a QSINE window function. All of the HRMAS spectra obtained on the 400 MHz spectrometer were calibrated using the DMSO peak, referenced to tetramethylsilane (TMS) at 0 ppm.

Artificial maturation was performed in gold tube reactors according to methods described by Al Darouich et al. (2006). **Constant temperature confined pyrolysis is performed at 200, 225, 250, 275 and 300°C during 9h.** Two tubes were used for each temperature/time condition, one for gaseous product analysis and the other for liquid product analysis. Gas analysis was performed on one gold tube pierced in a vacuum line equipped with a Toepler pump (Behar et al., 1989). Gas chromatography using a thermal conductivity detector was used to characterise and quantify of all the individual gases generated. Two liquid fractions were extracted : the first one recovered the hydrocarbons and the lightest NSOs compounds by pentane extraction, the second one was a dichloromethane extraction for the recovery of most of the heavy hydrocarbons and heteroatom-containing hydrocarbons. Extraction was performed by stirring under reflux for 1 hour. An initial aliquot of the pentane extract was used for quantification of the $C_6$-$C_{14}$ compounds. This aliquot was fractionated on a micro column of silica gel into saturated and aromatic compounds. An internal standard ($C_{25}$ *n*-alkane) was added for the quantification by GC-FID. The ***n*-$C_5$** compounds were not quantified because pentane is used as solvent. The second aliquot was evaporated and quantified. A second extraction was performed with dichloromethane (DCM) by stirring under reflux for 1 hour. The DCM extract was evaporated and quantified by weighing. The insoluble residue of this extraction was then dried and weighed. Mass and atomic balances were done in order to check the recovery yield of all the pyrolysis products.

## 4. Results and discussions

### *4.1. Initial structure*

To our knowledge, no molecular model of the Morwell lignite has been proposed to date in the literature. However a molecular model (Figure 1) of an Australian brown coal (or lignite B) from the Yallourn Open Cut (near the Morwell Open Cut, Holgate, 1985) was proposed by Hatcher (1989) from the experimental data of Bates and Hatcher (1989) and using as a structural motif the lignin model of fresh gymnosperm wood proposed by Alder (1977). This Adler model was transformed by applying experimental observations of peatification and early coalification reactions (Hatcher and Clifford, 1997).

We compared the atomic and structural compositions of this gymnosperm coal model with experimental data (elemental and solid-state $^{13}C$ NMR) obtained on the Morwell lignite (Behar and Hatcher, 1995) that is summarized in Table 1. Results show that, except for the amount of carboxylic groups, the distribution of oxygenated functional groups of the Morwell lignite is very similar to that of the gymnosperm coal model. The major difference is the relative proportion of aliphatic and aromatic carbons. The Morwell lignite is enriched in aliphatic carbon. The Morwell lignite, however, is an angiospermous wood as determined by the flash pyrolysis data shown in Behar and Hatcher (1995). In fact, angiospermous wood contains syringyl based units not found in gymnospermous wood (Philp et al., 1982). Therefore, use of the gymnospermous coal model of Hatcher (1989) is inappropriate to describe the structural nature of the Morwell lignite sample. Hence, we develop a new structural model for Morwell lignite that uses the same approach used by Hatcher (1989) but is derived from the angiospermous lignin structural motif of Nimz (1974) and is constrained by the new experimental data obtained in the current study (elemental analysis, FTIR, and NMR) along with reactions that have been proposed in previous studies (e.g., demethylation, removal of oxygen functional groups from side-chains, aryl ether cleavage-see Hatcher and Clifford, 1997).

Reactions of peatification and early coalification have been determined from observations of structural compositions of fresh wood samples and lignitic samples (Stout et al., 1988, Hatcher et al., 1989, Hatcher, 1989, Behar and Hatcher, 1995, McKinney and Hatcher, 1996). Figure 2 summarizes reactions we feel are important in transforming the carbon skeleton of the Nimz lignin model. In order to constrain reaction sites, distances between the reactant functional groups are computed from the coordinates of the individual atoms modeled in 3-D space as described below. Reactions are assumed to occur if the distances are below a distance of three single C-C bonds. Hence the lignin model shown in Figure 3 is constructed to represent the likely structure of the Morwell lignite built around the Nimz motif. We use Cerius$^2$ (version 4.8.1, Accerlys software) and minimized our structure with the UFF_VALBOND 1.1 force field (a combination of the original VALBOND method described by Root et al., 1993, augmented with non-orthogonal strength functions taken from Root, 1997 and the Universal Force Field of Rappé et al., 1992) to produce atomic coordinates.

With the reactions shown in Figure 2, the carbon skeleton of the Nimz model is rearranged, furan-like structures are removed and the proportion of side-chain hydroxyl/ether groups decreases significantly. Then the proportion of each functional groups are readjusted in the structure to match with quantitative NMR data (Table 2) and better represent the structure of coalified angiospermous wood that is equivalent in rank to the Morwell lignite . Thus, eleven hydroxyl functions are oxidized to carboxyl groups (reaction 6) and two are reduced to yield unsubstituted aliphatic carbons (reaction 7). Twenty methoxy functions are demethylated to yield hydroxyl aryl functions (reaction 8). Dehydroxylation of two hydroxyl aryl groups (reaction 9) allow for the adjustment of the proportions of aryl-O groups. The structural model thusly obtained for the Morwell coal is displayed in Figure 4. This model now serves as the basis for further reactions deduced from artificial maturation in sealed gold tubes. We modify this model in accordance with the chemical analyses of the residue and the liquid and gaseous products.

### *4.2. Observed chemical changes during thermal decomposition of Morwell lignite*

As mass balances given in Table 3 show, the total mass loss of the Morwell coal was 16.64% at 300 °C exposed for 9 h. The elemental composition changes significantly as demonstrated by the atomic composition of the recovered lignite that exhibits a precipitous decrease of both H/C and O/C ratios with values down to 0.69 and 0.22, respectively.

Concurrently, $CO_2$ is generated at temperatures starting as low as 200 °C, reaching a maximum yield of 12.99 wt% at the most severe conditions and representing 3.6% of the carbon balance. This value is similar to the combined carbon loss from the carboxyl (2.9%) and carbonyl (0.6%) functions as estimated by NMR (Figure 5).

#### *4.2.1. Structural evolution observed by DPMAS NMR.*

We calculate the carbon loss from NMR spectra in Figure 5 by multiplying area percentages for the various integration regions by the total residual carbon from the carbon mass balance that is based on a starting carbon content of 100 mg. Hence, in the structure, 17% of the carbonyl functions and 44% of the carboxylic groups are decomposed at 300 °C/9 h. The loss of methoxy groups (seen clearly in the NMR spectra of Figure 5) accounts for loss of 5.3% of carbons, but only 0.11 wt% of methane was recovered in gaseous pyrolysis products at the temperature of 300 °C. This lack of correspondence between loss of methoxy groups and methane generation suggests that almost all methoxy species lost are not converted into methane gas. There are multiple possibilities for the formation of end-products: one is that the methyl radical formed reacts with other components of the solid residue, another is reaction of the methyl radical with liquid or volatile products, another is reaction of the methyl radical with OH radical or water to form methanol which is difficult to measure by GC/MS and not included in the mass balance. Considering the uncertainty with which we understand the redistribution of the methyl group, we cannot specifically account for its fate at the moment, but we can suggest that methane is not the only end-product.

We cannot exclude the fact that part or the entire yield of the methane originates from pyrolytic degradation of aliphatic carbons in the Morwell coal. Butane, produced in higher yields than methane, is also likely to originate from pyrolytic degradation of aliphatic carbons in the Morwell coal. In fact, we calculate that a combined 2.76 wt% of methane and butane are produced, with no contributions from other volatile hydrocarbon gases. The NMR data shows that 8.22% of aliphatic carbons (CH, $CH_2$, and $CH_3$ not including O-substituted aliphatic groups) are lost from the structure of the Morwell coal heated to 300 °C, more than enough to account for the volatile hydrocarbons produced, including methane, if one assumes that aliphatic side chains are the source of these volatile species.

If we include O-substituted aliphatic groups and carboxyl/carbonyl species in our carbon balance, there is a substantial loss of lignin side-chain carbons in the Morwell coal considering that the initial sample contained 29% C of total carbons associated with such side chains and the residue contains only 16%. Thus, the carbon skeleton released half of its aliphatic side chains by defunctionalisation and/or depolymerisation and/or pyrolytic degradation.

Interestingly, the total amount of aromatic carbons (*fa* = aroC + aryl-O carbons) increased by 2% (from 64.9% to 66.9% of the carbon). The uncertainty of the peak area in solid-state NMR is estimated to 3% corresponding to a range of uncertainty for the total aromatic carbon of more or less 2%. This suggests that no significant amount of aromatic units is released from the initial structure upon thermal stress. Behar and Hatcher (1995) show that $C_{6+}$ pyrolyzate, extract from residue obtained between 200 and 300°C, is essentially composed of aromatics structures (benzene, phenol, catechol, guaiacol, synringol and naphthalene were identified). Thus, if such structures are formed, they should be recovered in the $C_{6+}$ pyrolyzate. In fact, we measure during the experiment at 300°C during 9h, as low as 0.90wt % of $C_{6+}$ pyrolyzate. Using the atomic composition of the extracts (pentane and DCM $C_{14+}$ extract) we can estimate that 0.69% of aromatic carbons are lost which is not significant compared to the uncertainty in area measurements of 2%. This confirms that the amount of aromatic carbon, in the residue recovered at 300°C, is similar to the amount in the initial sample. However, the NMR spectra in Figure 5 show significant changes to the aromatic region. The aryl-O disappears during artificial maturation and the amount that disappears (6.5% of the total carbon) is not exactly equivalent to the amount of aromatic carbon that appears (8.6%) suggesting that aryl-O carbons are transformed to aromatic carbons not bearing an O and that additional aliphatics carbons are converted to aromatic carbons. The proportion of aryl-O functions initially represents 39.8% of total aromatic carbons (*fa*) and these decreases to 28.8% in the recovered residue at 300 °C. Hence, we calculate that 6.55% of carbons correspond to 25% of the aryl functions that are converted into aromatic carbons by losing an oxygen substituent. Two processes may explain the conversion of the aryl-O carbons to aromatic carbons: Behar and Hatcher (1995) demonstrated that two dihydroxyle rings link by an ether group is decomposed to one dihydroxyle units and one monohydroxyle units, another process is that the aryl functions lose an hydroxyl group like by reaction 9.

Considering that the total loses of aryl function is decomposed by deshydroxylation of aryl-OH functions forming water and that lost of hydroxyl groups in the lignin side chain generated water, the maximum of carbon involving in dehydration of the mature sample is estimated to 8.2% C (6.5 % of aryl-O carbons + 1.7% of alkoxyl carbons) during artificial maturation.

#### *4.2.2. Structural evolution observed by FTIR spectroscopy*

In agreement with the solid state NMR, the infrared spectra Figure 6 show a sharp decrease of the C=O and C-O bands relative to the defunctionalisation of the side chain. Only vibration bands of the aromatic skeletons and of OH are observed in the spectra at 300°C. Because of conversion of aryl-O carbons to aromatic carbon, noticed by quantitative NMR, bi- (guaiacol untis) and trihydroxyl (synringol) rings are converted in bi- (guaiacol) and monohydroxyl (p-hydroxyphenol) rings suggested for instance in Hatcher et al. (1989). Asymmetric deformation bands at 1470 $cm^{-1}$of the aromatic skeleton are replaced by symmetric deformation bands at 1370 $cm^{-1}$, in the infrared spectra. This change in vibration band is in agreement with the convertion of the synringol and guaiacol asymmetric units into symmetric units of p-hydroxyphenol.

### *4.2.3. Structural evolution observed by HRMAS NMR*

The HRMAS data shown in Figure 7 and 8 for both the unreacted Morwell coal and the residue at 300 C/9h provide additional clues as to the specific transformations that occur during artificial maturation, except for the aromatic region. In the aromatic region of the HSQC spectra (Figure 7 and 8) the cross peaks for both the untreated and matured samples are dispersed over a wide chemical shift range and it is almost impossible to assign any one cross peak to a specific structure. It is apparent that the number of cross peaks diminishes with increasing maturation, and this may be due to the fact that aromatic rings are becoming deprotonated as the result of heating. Once deprotonated the cross peaks for aromatic carbons disappear. On the DPMAS spectra, depicted as a projection on the ordinate, the amount of aromatic carbon protonated represent a small amount of the total area ascribed to aromatic carbons. It is important to underline that those small amount do not represent the precise amount of aromatic carbons protonated but the minimum of those protonated carbons detected by HSQC NMR.

In HRMAS spectra quaternary carbons, such as ketone and carboxyl carbons, are not detected, thus only the aldehyde, hydroxyl and methoxy oxygenated functional groups are observed (Figure 7). Aldehyde functional groups, identified as 1, are observed in both TOCSY spectra (Figure 9) of the untreated and the mature sample which show that aldehydes persist upon thermal decomposition at 300 °C. However those aldehydes are not be present in large amounts, because no signal is detected in the region of the carbonyl groups in the HSQC spectra despite the fact that around 3% are quantified by solid state NMR in the initial coal and the residue recovered at 300°C. This suggests that carbonyl groups are mainly ketones which do not show signals in HSQC or TOCSY spectra.

In standard lignin structural units, the hydroxyl functions may be substituted in the α, β or γ positions of the aliphatic side chain (see Figure 1b). Though, in the initial structure only hydroxyl functions in the γ position are observed (cross peak 5 in Figure 7, and 8). We cannot exclude the fact that, α and β hydroxyl functions fall below the detection threshold. In fact, 0.4% of hydroxyl functions are quantified in the DPMAS spectrum of the recovered residue whereas in the 2D spectra (HSQC and TOCSY) no signal is observed for these structures. As observed in quantitative NMR spectrum, the intensity of methoxy functional groups assigned to signal 6 is significantly decreased compared to the original coal sample.

Aliphatic carbons (Figure 8) are assigned to regions 7, 8, 9 corresponding to –CH-, -$CH_2$- and -$CH_3$. In region 7, additional signals (7B, 7C) appear in the residue recovered at 300°C and signal 7A, ascribed to -CH- in benzylic positions, is detected in both the initial sample and the heated residue. The increase of aliphatic -CH- groups in the heated residue is probably associated with increased cross linking of the aliphatic side chains. In region 8 of the unheated sample, cross peaks are dispersed and of low intensity, whereas, in the mature sample, cross peaks are well defined, more intense, and less abundant. This is consistent with a process involving defunctionalisation of the aliphatic side chains, a process described previously for the early coalification process (Hatcher and Clifford, 1997; Solomon et al., 1988). In the unheated sample containing mainly lignin structural units, aliphatic carbons (-CH-, -$CH_2$- and -$CH_3$) are typically adjacent to carbons substituted by carboxyl, carbonyl and hydroxyl groups. This can explain why their chemical shifts are broadly dispersed in the spectrum. Upon thermal stress, oxygenated functional groups are released, leading to an increasing signal strength for aliphatic carbons and to more uniform structural characteristics which translate to fewer peaks. Similarly, the TOCSY spectrum (Figure 9 and 10) of the unheated sample contains more dispersed aliphatic cross peaks than the heated sample confirming that aliphatic carbons are less diverse in structural makeup following the artificial maturation. Cross peak 8A is assigned to benzylic $CH_2$ on the aliphatic side chains and carbons 8B and 8C are attributed to $CH_2$ groups that are β or γ to the aromatic carbons. The $^1H$ chemical shifts of peaks in region 8B are more downfield than the $^1H$ chemical shifts of peaks 8C suggesting that 8B structures are more proximal to aromatic rings or to oxygenated functions than are 8C structures. Increased heating leads to a shift in peaks of region 8B to lower $^1H$ chemical shift values. Defunctionalisation of the side chain is probably responsible for this change. The transformation of structures associated with 8D (unheated sample) to 8D' (heated sample) is attributed to a rearrangement of the side chain carbons to form a 7-member alicyclic structure as shown in Figure 8, structure V. The broadening of cross peak 8D' is consistent with the presence of the naphthenic structures linked to aromatic rings as shown in Figure 8, structures IX and X. The aromatic naphthenes are also consistent with the evolution of vibrations bands in the infrared spectrum of the heated sample (see discussion above). An increased abundance of methyl groups (region 9) is observed in the spectra of the residue recovered at 300°C. Signals associated with box 9A' correspond to methyl groups in terminal positions on the propyl or ethyl side chains. Peaks in box 9B are assigned to methyl groups in benzylic positions, an indication that part of the original aliphatic lignin side chains has been cleaved.

The TOCSY spectra are shown in Figures 9 and 10 and a table of spectral assignments is also given in Figure 10 for protons in the aliphatic region. The information obtained from these spectra is entirely consistent with what is observed in the HSQC spectra; except that a peak is observed for aldehyde protons (1') in the heated coal (Figure 9).This is due to the higher sensitivity of the TOCSY than the HSQC method. In the aromatic region (Figure 9), $^1H$-$^1H$ couplings decrease because the amount of protonated aromatic carbons decreases in the structure (see discussion of HSQC NMR data above). In the aliphatic region (Figure 10), only $CH_3$-$CH_2$ and $CH_2$-$CH_2$ couplings are detected in both spectra. The absence of peaks $J_B$ and $J_D$ in the spectrum of the initial sample shows that the amount of coupling through two and three bonds is very low. This confirms that CH, $CH_2$ and $CH_3$ are widely dispersed in the structure and that carbons substituted by oxygenated functional groups disrupt long range connectivity within a single spin system. The cross peak $J_H$, appearing only in the thermally stressed sample, is assigned to ethyl side chains on aromatic rings which is in agreement with the DPMAS NMR data showing that pyrolytic degradation of the aliphatic side chains occurs during thermal stress.

## 5. Overall processes of maturations

**The changes in molecular-level composition are globally quantified by elemental analysis and solid state NMR, providing constraints for the chemical transformation of the lignin model to the rank of lignite. In addition to gaseous and liquids products derived from artificial maturation, structural evolution of the coalified wood model is described. The carbon balance and thermal decomposition processes are summarized in Figure 11. Pyrolysis experiments, performed at 300°C over a period of 9h, convert 10.6% of the Morwell coal carbon into gaseous and liquid products. Early transformations involve mainly the rearrangement of the coal, and generation of gaseous products as described previously by Solomon et al. (1988) as well as Behar and Hatcher (1995). First, a large amount of $CO_2$ is generated; then, gaseous hydrocarbons are produced in lower amounts. As much $CO_2$ (5.8% of carbon) is released by defunctionalisation as is generated as gaseous and liquid hydrocarbons (4.8% totally as methane, butane and $C_{14+}$ liquid hydrocarbons). NMR data confirm that $CO_2$ is formed by defunctionalisation of carboxyl and carbonyl functional groups. The methane generated is insufficient to be associated with the loss of methoxy groups in NMR spectra (5.3% of carbons). At low maturation levels, methyl radicals, presumably**

**derived from removal of methoxy groups, may be involved in multiple reaction pathways forming various pyrolytic products as well as solid, liquid or gaseous hydrocarbons. Alternatively, methoxy groups may form methanol or formaldehyde, both of which could not be measured directly in this study. Methane and butane could derive from reactions of methyl radicals but they also could evolve from pyrolytic degradation of the aliphatic side chains in lignin. This process could involve defunctionalisation of the side chain followed by pyrolysis.**

**Thermal evolution of the insoluble portion of the coal leads to an increase in cross linking and in the presence of symmetric aromatic structures. This process is in complete agreement with the cross-link processes at low temperature proposed by Solomon et al. (1990) who suggested that, at low temperature and prior to tar evolution, cross linking of the kerogen is correlated to $CO_2$ loss, water and light hydrocarbons generation. We determined by quantitative NMR that half of the side chains disappeared by formation of gaseous products ($CO_2$, methane and butane).**

**Behar and Hatcher (1995) have shown that the liquid $C_{14+}$ fraction contains aromatic structural units. However, these are minor as our carbon balance indicates that they represent only 1.1% of the carbon. This suggests that aromatic rings in the residue are not significantly lost. Characterization by HRMAS NMR of the mature sample shows that linkages between the aromatic rings and the aliphatic side chains increase but do not enhance the aromaticity of the structure. Oxygenated and protonated carbons on the aromatic rings are converted to carbon-carbon linkages and the proportion of CH increases in the side chains. The proportion of naphthenic rings seems to increase upon maturation, perhaps because of the alteration of the linear side chains.**

## 6. Conclusions

This paper is part of a study that seeks to define the relative importance of defunctionalisation and cracking processes during early thermal decomposition of fossil organic matter from numerous sources. Understanding and quantifying those processes is paramount to developing improved kinetic models that are used to evaluate the extent of petroleum generation in sedimentary basins. This paper examines the early reactions for the thermal evolution of a Type III kerogen using as a starting point the chemistry of a sample derived from coalified wood collected from the Gippsland Formation, Morwell open cut mine in Victoria, Australia. The experimental data obtained from artificial maturation in a closed-system reactor will be used for comparison with results of a joint study that proposes use of a new technique for determining maturation changes, that of molecular dynamics simulations with a reactive force field. The structural model for the Morwell coal proposed in this paper (Figure 4) is the only input data of the molecular dynamics simulations. For this reason, the lignin model of Nimz, (1974) is selected to represent the angiospermous origin of the Morwell sample.

All the structural changes mentioned in this paper lead to a rearrangement of the coal that can be described through a molecular model that is the subject of a future paper. We expect in this future study to reproduce by molecular dynamic simulations the chemical processes experimentally observed in the current paper in order to validate the simulation procedure and to confirm, from a theoretical point of view, reactions processes proposed here and in the literature. In a way, positive results from the dynamic simulations will also validate the structural model for the Morwell coal.

## Acknowledgements

The experimental facilities used for this research were provided by the IFP (Institut Français du pétrole), and the College of Sciences Major Instrumentation Cluster in Old Dominion University. Fellowship support was provided by the ANRT (Association Nationale de la Recherche Technique), CIFRE grant #458/2004.

The authors would like to thanks Isaiah Ruhl for assistance with the NMR experiments. We also thank Neal Gupta (Department of EAPS- MIT) and anonymous reviewer for their constructive review of this manuscript.

**Table captions**

**Table 1.** Comparison of the structural composition of the Morwell sample (Behar and Hatcher 1995) with that of the brown coal model of Hatcher (**1989**).

**Table 2.** Structural composition of the fresh lignin model (Nimz, 1974) and of the model and sample of the Morwell coalified wood.

**Table 3.** Mass and atomic balances of the experimental thermal decomposition of the Morwell coal.

## Figure captions

**Figure 1.** (a) Brown coal molecular model from Hatcher (1990). (b) Typical units in lignin : numbering convention

**Figure 2.** Early diagenesis reactions selected to transform the lignin structure to mature rank of lignite. (Hatcher, 1989; Hatcher and Clifford, 1997; Payne and Ortoleva, 2002)

**Figure 3.** Beech lignin model from Nimz (1974)

**Figure 4.** Structural model of the Morwell sample at coal rank of lignite.

**Figure 5.** DPMAS $^{13}C$ NMR data of the initial sample of Morwell lignite (A ) and residue (B) recovered after thermal stress (300°C/9h). The inset table provides quantitative measurements of the relative contributions of the various carbons. The % loss of carbon during artificial maturation is also indicated. The errors (±) are given for each calculated value and represent a relative error of 3% for peak area measurements.

**Figure 6.** FTIR spectra of the initial sample of Morwell lignite (A) and the residue (B) recovered after thermal stress (300 °C/9 h). Various assignments for stretching ($\nu$) and deformation ($\delta$) frequencies.

**Figure 7.** HSQC spectra of the initial sample of Morwell lignite and the residue recovered after thermal stress (300 °C/9 h). Boxed out regions are discussed in the text. The solvent peak is for DMSO. The left ordinate projection is the respective DPMAS $^{13}C$ NMR spectrum.

**Figure 8.** Extended aliphatic region of the HSQC NMR spectra in Figure 9 of the initial Morwell lignite and the residue recovered at 300°C/9 h. Structural assignments for the indicated carbon are presented in a table below of the spectra.

**Figure 9.** TOCSY spectra of the initial Morwell lignite and the residue recovered at 300°C/9 h. Structural assignments for some cross peaks are listed in Figure 8. Boxed out regions are discussed in the text. The solvent peak is for DMSO.

**Figure 10.** Expanded view of the TOCSY spectrum in Figure 9. Structural assignments for the $^{1}Hs$ are presented in a table below of the data with chemical shifts indicated for each coupled system.

**Figure 11.** Carbon mass balance of the Morwell lignite sample during 300°C/9h closed pyrolysis.

**Table 1.** Comparison of the structural composition of the Morwell sample (Behar and Hatcher 1995) with that of the brown coal model of Hatcher (**1989**).

| chemical type | Morwell Sample (Behar et al. 1995) | Brown coal Model (Hatcher 1989) |
|---|---|---|
| | atomic C (%) | |
| C=O | 2.90 | 2.78 |
| COOH | 4.00 | 1.85 |
| $C_{aro}$-$OR_1$ | 22.80 | 21.30 |
| $C_{aro}$- | 39.50 | 45.37 |
| O-$CH_3$ | 3.10 | 4.63 |
| C-O-$R_2$ | 4.00 | 5.56 |
| -CH, -$CH_2$, -$CH_3$ | 23.10 | 18.52 |
| total | 99.40 | 100.01 |
| $f_a$ | 62.30 | 66.67 |
| formula | - | $C_{108}H_{102}O_{34}$ |
| H/C | 0.89 | 0.94 |
| O/C | 0.33 | 0.31 |

*fa*: aromaticity, $R_1$ : H or -$CH_3$, $R_2$: H, -C-

**Table 2.** Structural composition of the fresh lignin model (Nimz, 1974) and of the model and sample of the Morwell coalified wood.

| chemical structure | Lignin Model | Morwell Model | Morwell Sample |
|---|---|---|---|
| | % C | % C | % C |
| -C=O | 3.1 | 3.0 | 3.5 |
| -COOH | 0.4 | 6.4 | 6.6 |
| $-C_{aro}-OR_1$ | 22.8 | 24.4 | 25.8 |
| $=C_{aro}-$ | 35.1 | 39.7 | 39.1 |
| $-C-O-R_2$ | 18.9 | 3.0 | 2.1 |
| $-O-CH_3$ | 13.1 | 6.0 | 6.1 |
| $-CH, -CH_2, -CH_3$ | 6.6 | 17.5 | 16.9 |
| total | 100.0 | 100.0 | 100.0 |
| *fa* | 57.9 | 64.1 | 64.9 |
| Formula | $C_{259}H_{306}O_{93}$ | $C_{234}H_{216}O_{96}$ | |
| H/C | 1.18 | 0.92 | 0.94 |
| O/C | 0.359 | 0.410 | 0.390 |

*fa*: aromaticity, $R_1$ : H or $-CH_3$, $R_2$: H, -C-

**Table 3.** Mass and atomic balances of the experimental thermal decomposition of the Morwell coal.

| T | t | $CO_2$ | $CH_4$ | $C_4$ | $C_7$-$C_{14}$ | $C_{14+}$ | | | Residue | | |
|---|---|---|---|---|---|---|---|---|---|---|---|
| | | | | | | n-$C_5$ | DCM | Total | yield | H/C | O/C |
| °C | h | mg/g | | | | | | | mg/g | atomic ratio | |
| initial | | | | | | | | | | 0.94 | 0.39 |
| 200 | 9 | 45.8 | < 0.1 | 2.5 | < 0.1 | 1.3 | 2.6 | 52.2 | 947.8 | 0.88 | 0.34 |
| 225 | 9 | 66.5 | < 0.1 | 4.6 | < 0.1 | 1.4 | 4.0 | 76.5 | 923.5 | 0.86 | 0.31 |
| 250 | 9 | 93.2 | 0.1 | 5.5 | < 0.1 | 1.4 | 6.5 | 106.7 | 893.3 | 0.80 | 0.29 |
| 275 | 9 | 105.7 | 0.4 | 8.7 | < 0.1 | 1.4 | 6.2 | 122.4 | 877.6 | 0.73 | 0.26 |
| 300 | 9 | 129.9 | 1.1 | 26.5 | < 0.1 | 1.7 | 7.2 | 166.4 | 833.6 | 0.69 | 0.22 |

**Figure 1.** (a) Brown coal molecular model from Hatcher (1990). (b) Typical units in lignin : numbering convention

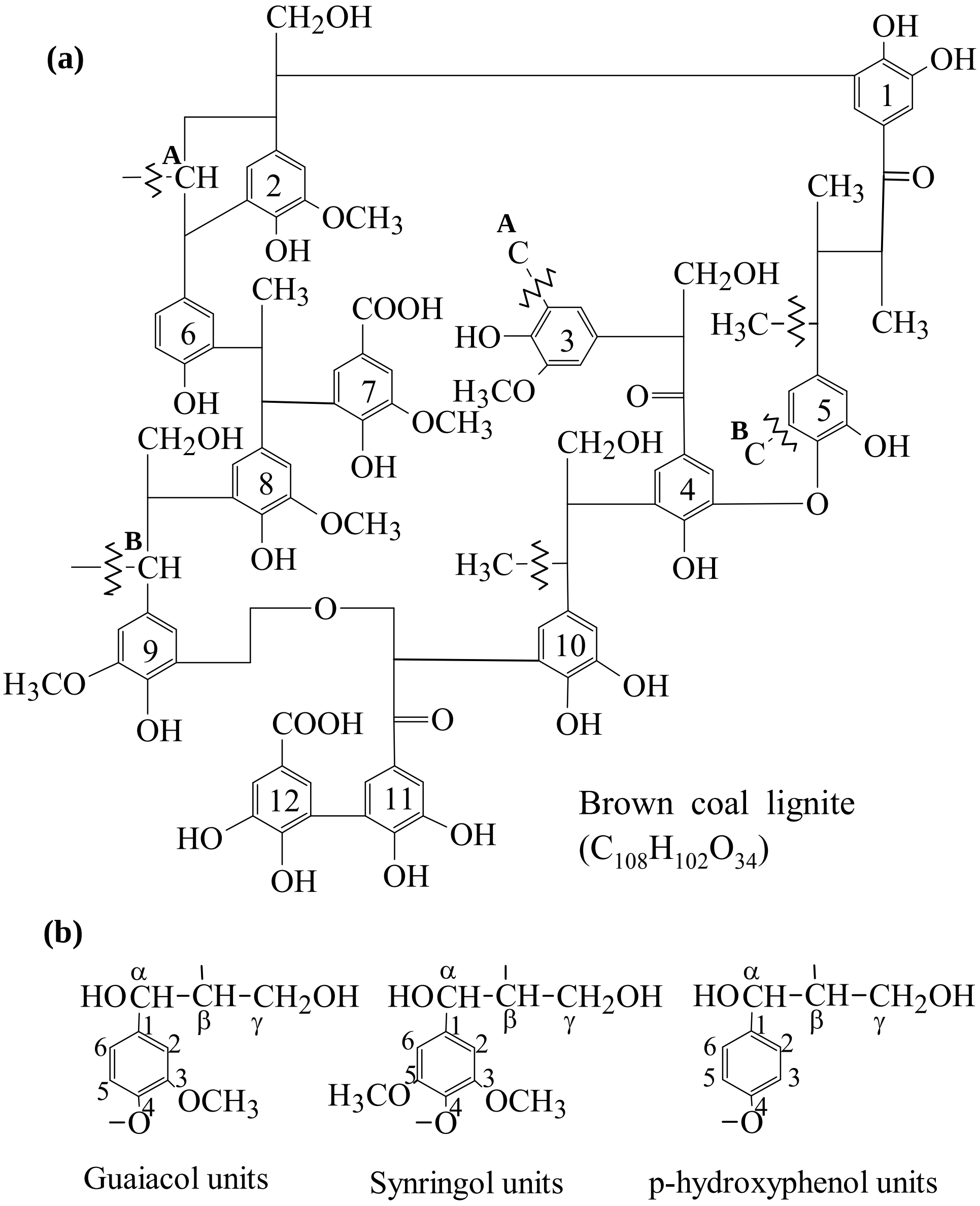

**Figure 2.** Early diagenesis reactions selected to transform the lignin structure to mature rank of lignite. (Hatcher, 1989; Hatcher and Clifford, 1997; Payne and Ortoleva, 2002)

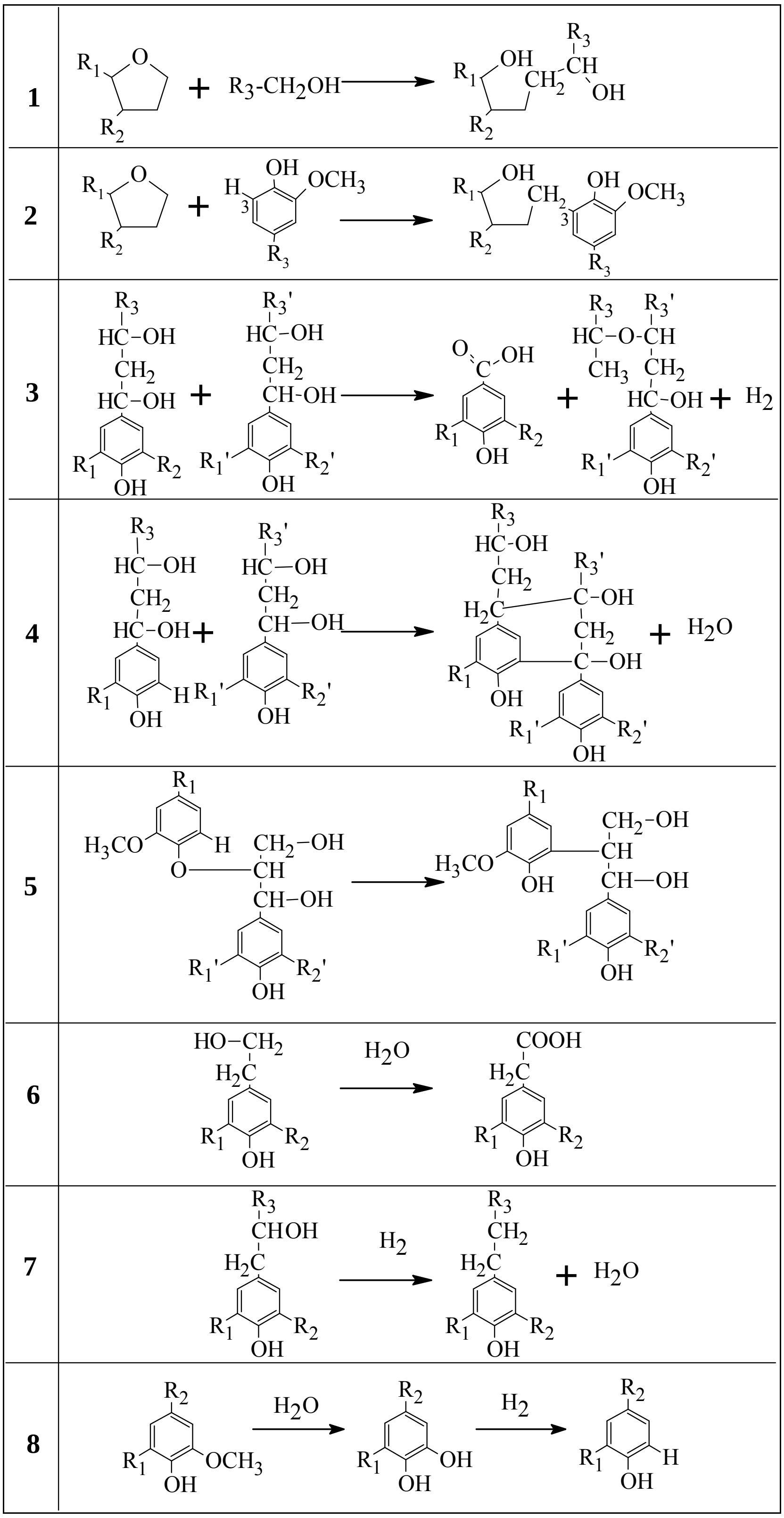

**Figure 3.** Beech lignin model from Nimz (1974)

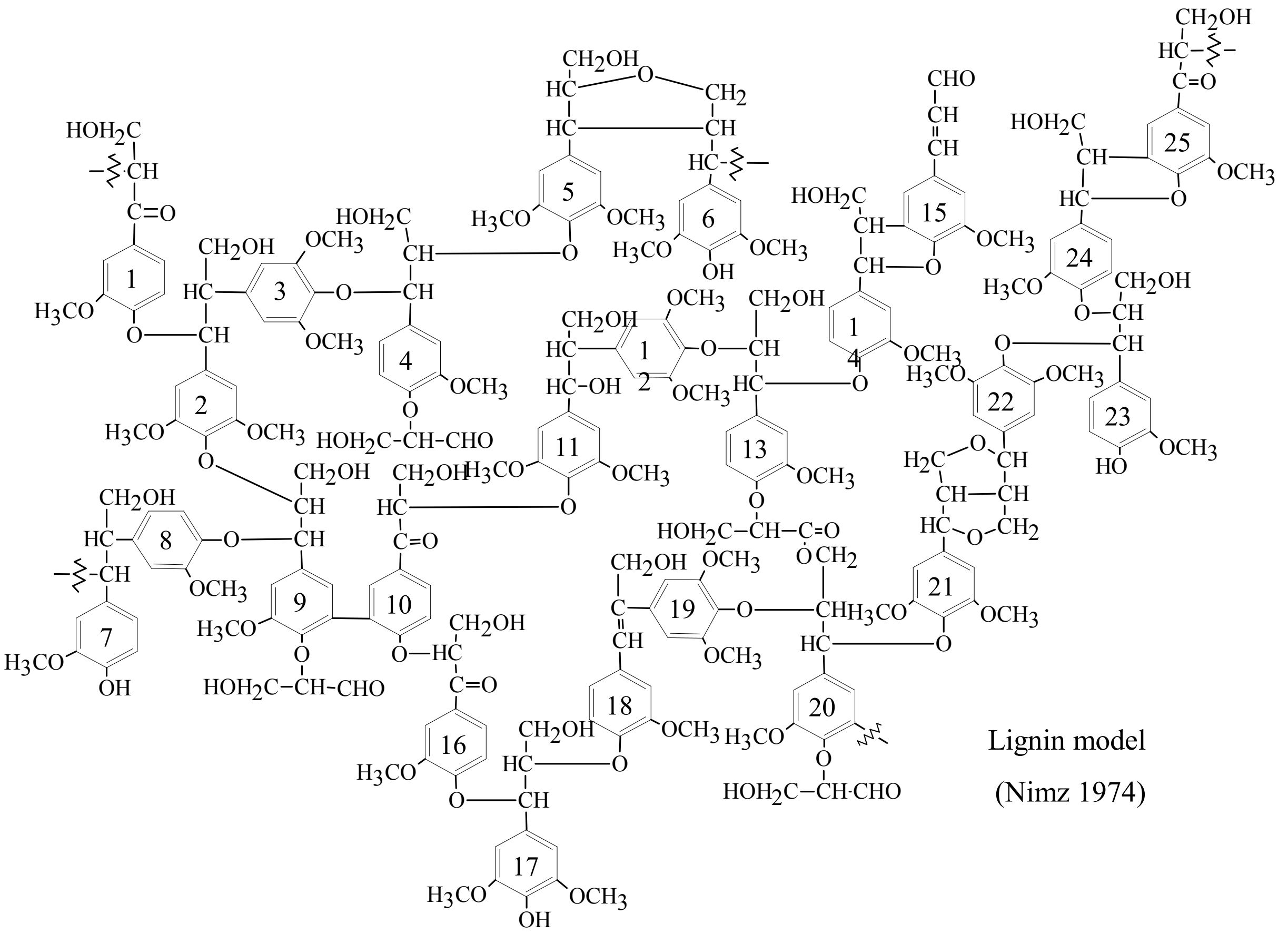

**Figure 4.** Structural model of the Morwell sample at coal rank of lignite.

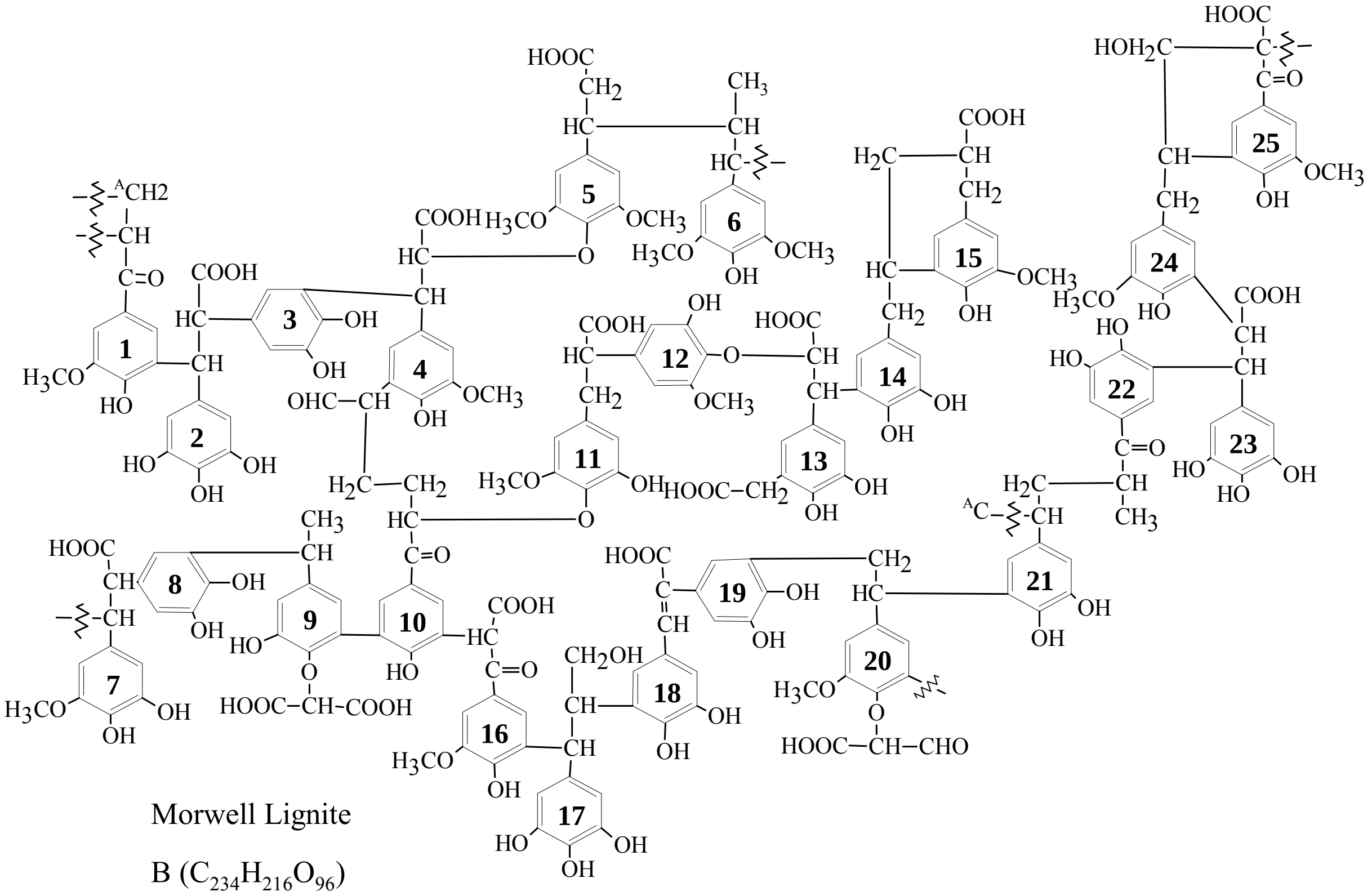

Morwell Lignite

B ($C_{234}H_{216}O_{96}$)

**Figure 5.** DPMAS $^{13}C$ NMR data of the initial sample of Morwell lignite (A ) and residue (B) recovered after thermal stress (300 °C/9 h). The inset table provides quantitative measurements of the relative contributions of the various carbons. The % loss of carbon during artificial maturation is also indicated. The errors (±) are given for each calculated value and represent a relative error of 3% for peak area measurements.

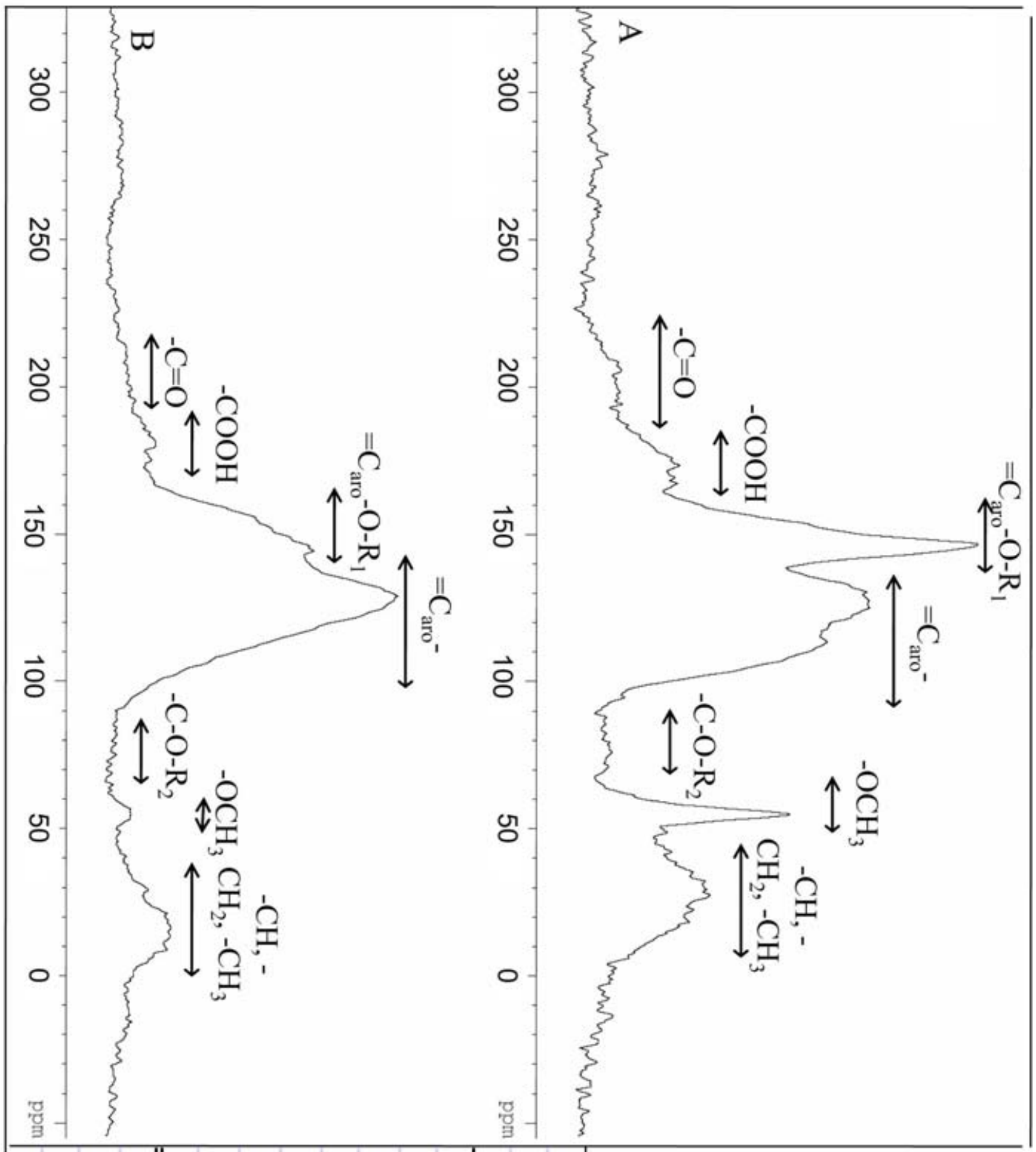

| chemical type | Initial sample | 300°C 9h | total C loss % |
|---|---|---|---|
| | Carbon (atomic %) | | |
| -C=O | 3.5 ± 0.1 | 2.9 ± 0.1 | -0.6 |
| -COOH | 6.6 ± 0.2 | 3.7 ± 0.1 | -2.9 |
| =Caro-O-R$_1$ | 25.8 ± 0.8 | 19.2 ± 0.6 | -6.5 |
| =Caro- | 39.1 ± 1.2 | 47.7 ± 1.4 | 8.6 |
| -C-O-R$_2$ | 2.1 ± 0.06 | 0.4 ± 0.01 | -1.7 |
| -O-CH$_3$ | 6.1 ± 0.2 | 0.8 ± 0.02 | -5.3 |
| -CH, -CH$_2$-, -CH$_3$ | 16.9 ± 0.5 | 8.7 ± 0.3 | -8.2 |
| total | 100.0 | 83.4 | -16.6 |
| *fa* | 64.9 ± 2 | 66.9 ± 2 | 2.1 |
| %( =Caro-O-R$_1$) | 39.8% | 28.8% | -11.0% |
| %( =Caro-) | 60.2% | 71.2% | 11.0% |
| ali/aro | 0.26 | 0.13 | |

**Figure 6.** FTIR spectra of the initial sample of Morwell lignite (A) and the residue (B) recovered after thermal stress (300°C/9h). Various assignments for stretching (ν) and deformation (δ) frequencies.

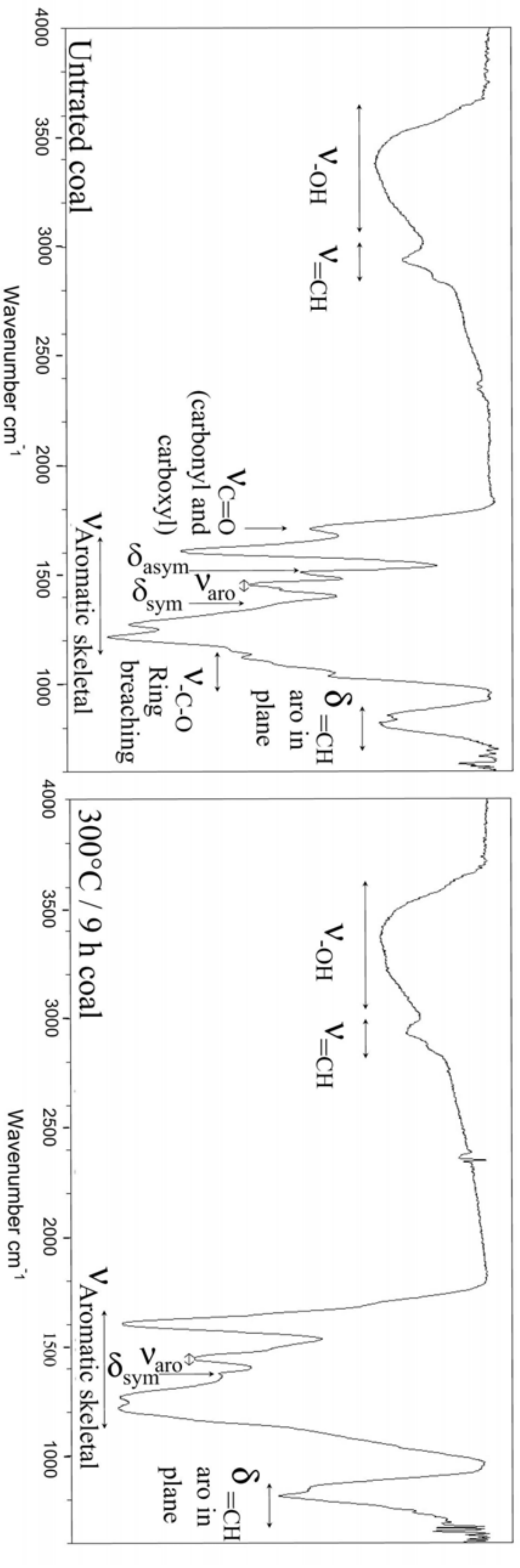

**Figure 7.** HSQC spectra of the initial sample of Morwell lignite and the residue recovered after thermal stress (300 °C/9 h). Boxed out regions are discussed in the text. The solvent peak is for DMSO. The left ordinate projection is the respective DPMAS $^{13}C$ NMR spectrum.

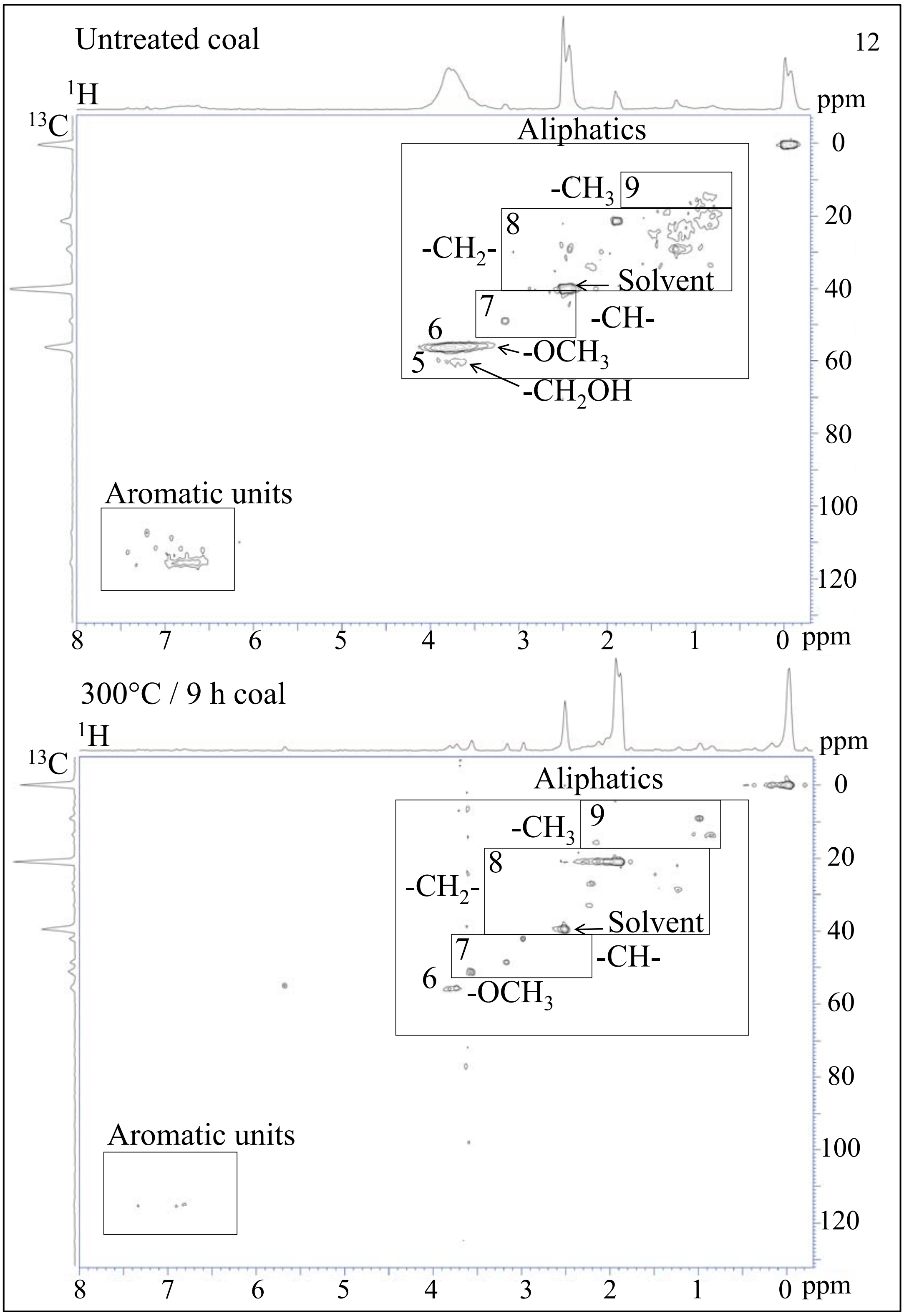

**Figure 8.** Extended aliphatic region of the HSQC NMR spectra in Figure 9 of the initial Morwell lignite and the residue recovered at 300°C/9 h. Structural assignments for the indicated carbon are presented in a table below of the spectra.

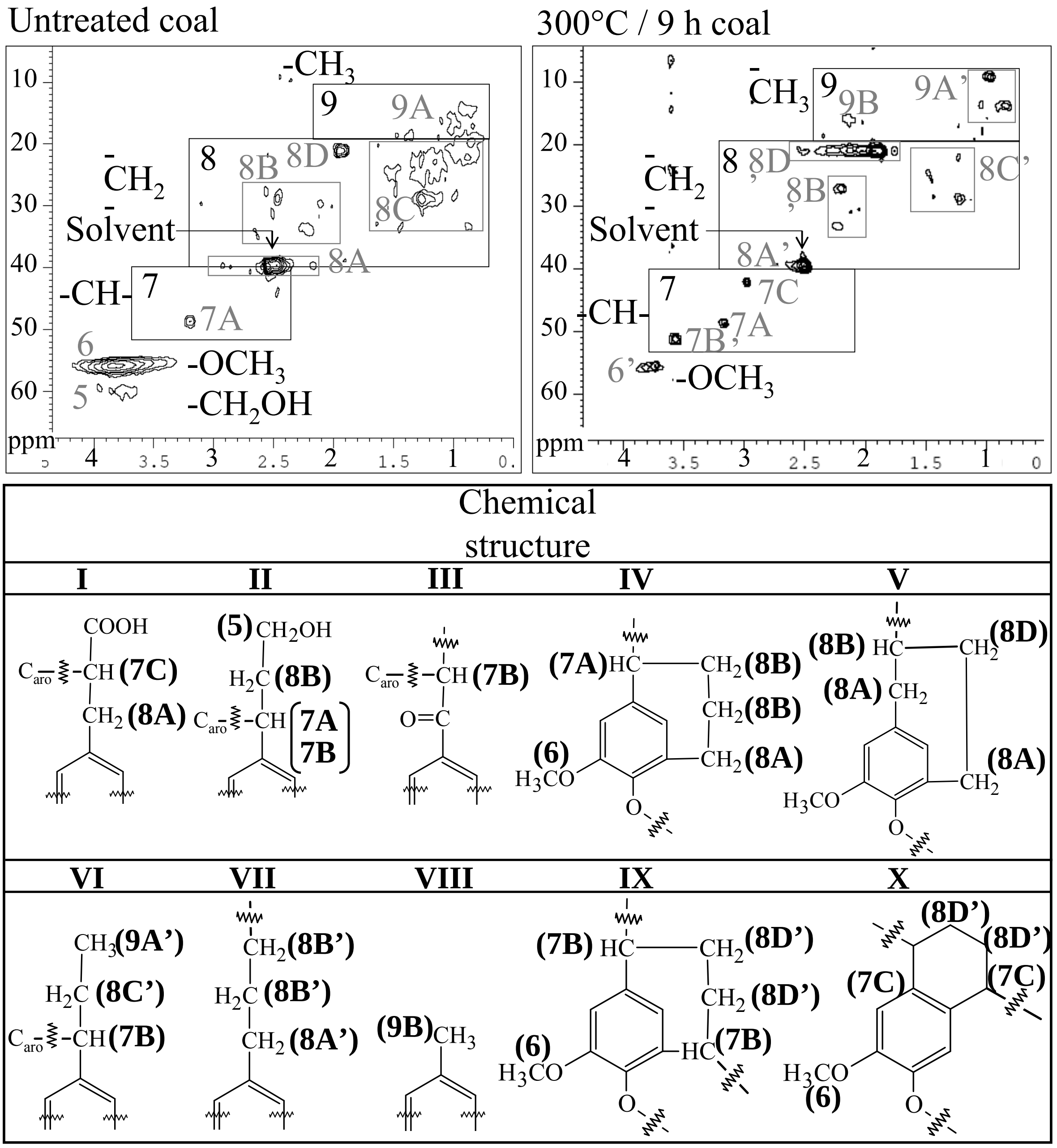

**Figure 9.** TOCSY spectra of the initial Morwell lignite and the residue recovered at 300°C/9 h. Structural assignments for some cross peaks are listed in Figure 8. Boxed out regions are discussed in the text. The solvent peak is for DMSO.

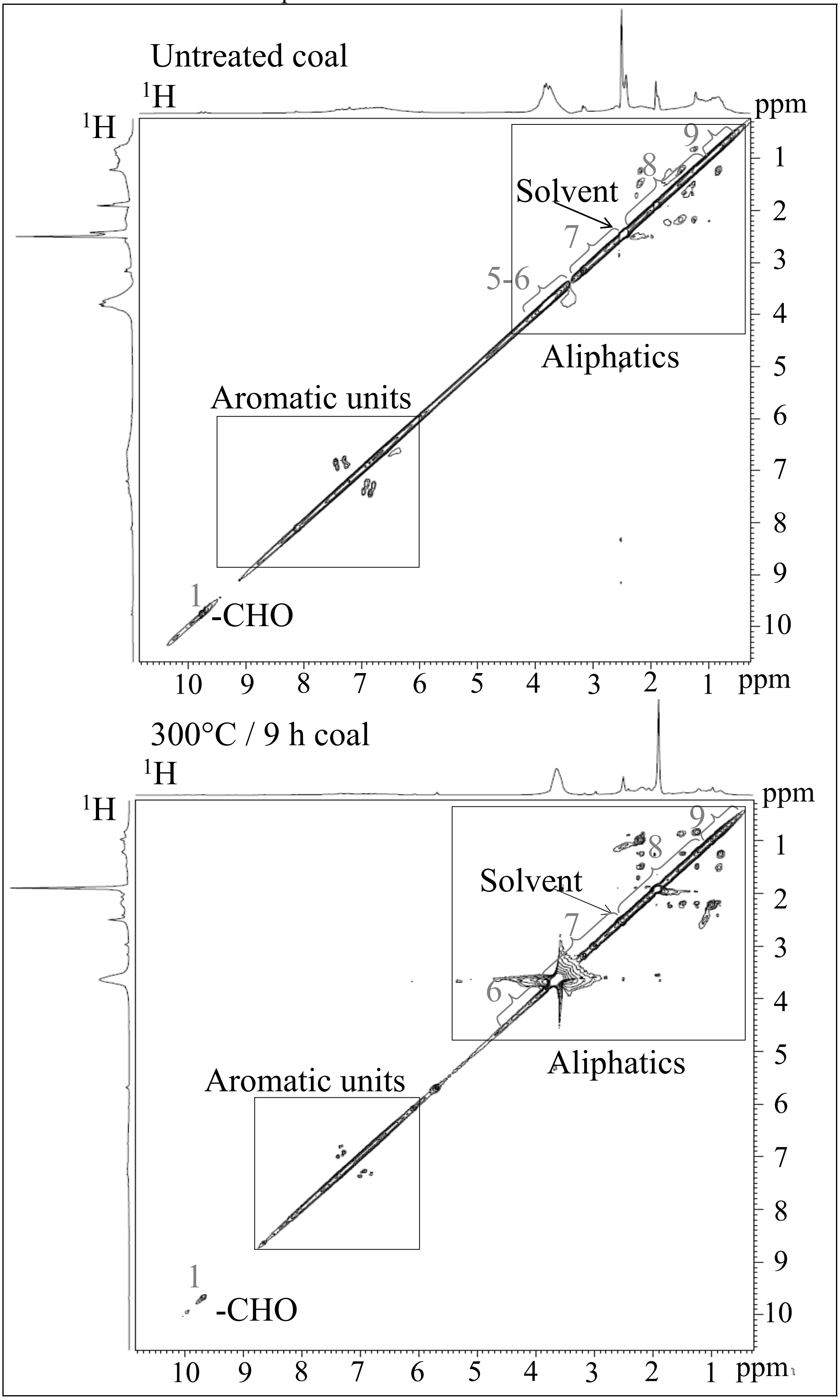

**Figure 10:** Expanded view of the TOCSY spectrum in Figure 9. Structural assignments for the [1]Hs are presented in a table below of the data with chemical shifts indicated for each coupled system.

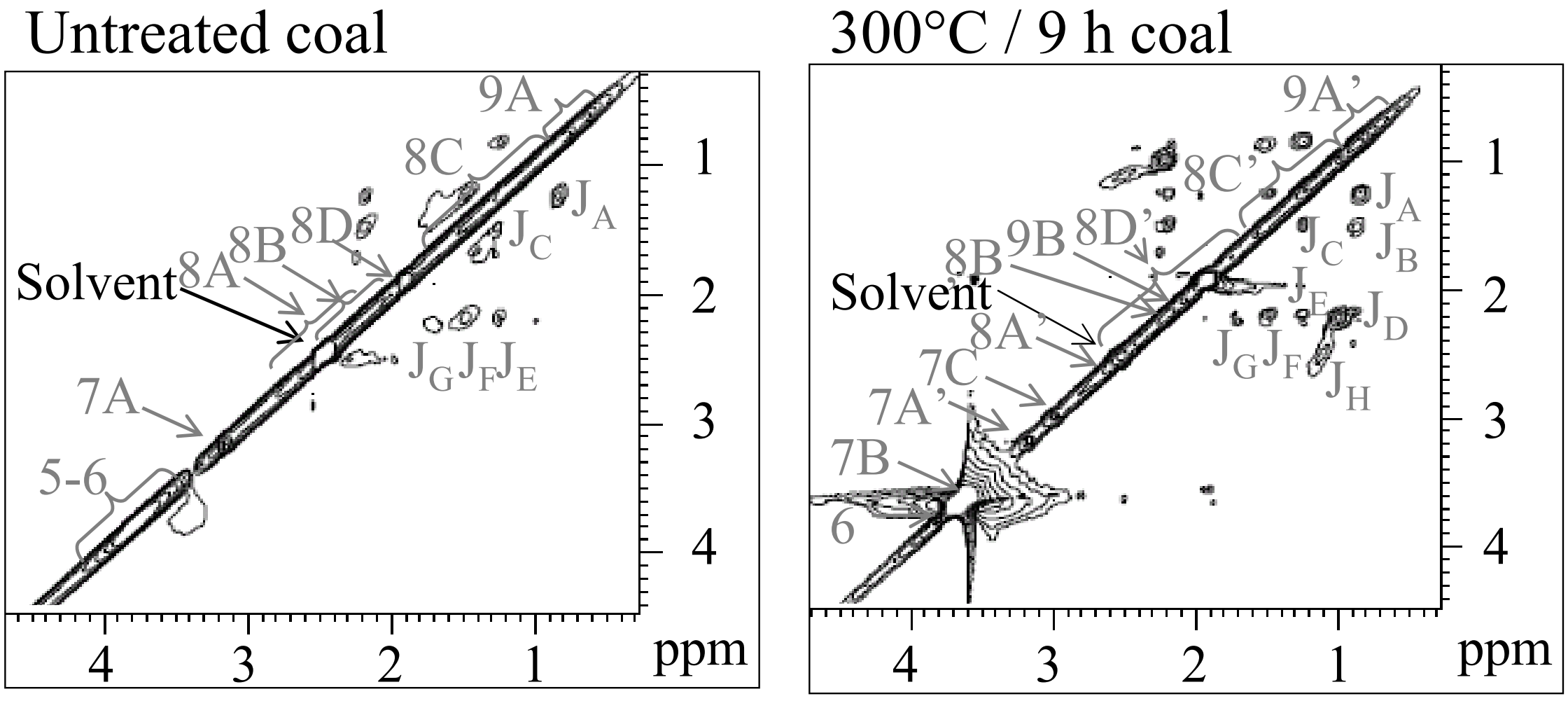

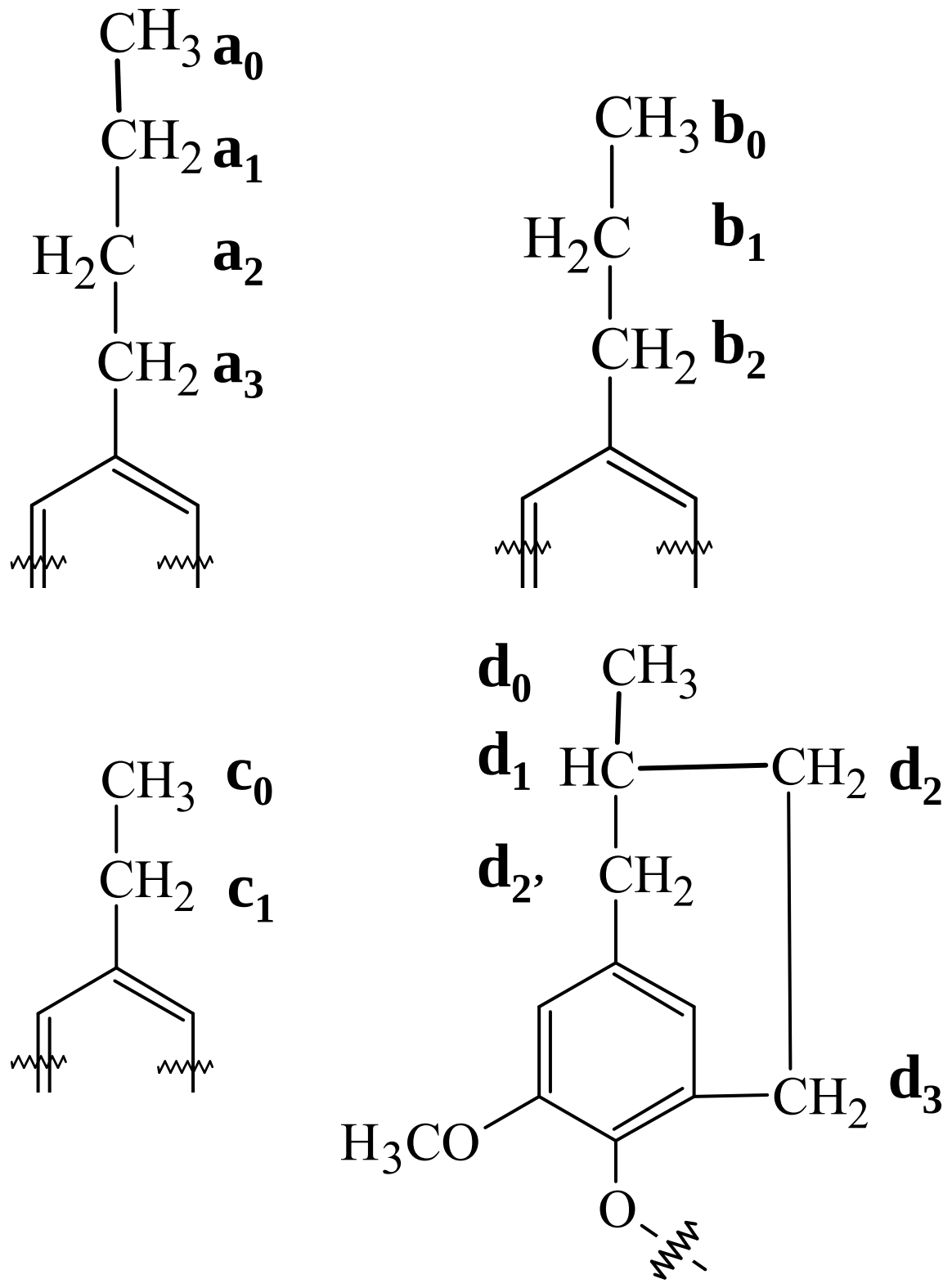

| Assignment: coupling | Assignment: carbons | $\delta^1H$ $H_x$ (ppm) | $\delta^1H$ $H_y$ (ppm) |
|---|---|---|---|
| $J_A$ | $a_0$-$a_1$, $b_0$-$b_1$, $d_0$-$d_2$ | 0.8 | 1.3 |
| $J_B$ | $a_0$-$a_2$, $b_0$-$b_2$, $d_0$-$d_1$ | 0.8 | 1.5 |
| $J_C$ | $a_1$-$a_2$, $b_1$-$b_2$, $d_1$-$d_2$ | 1.2 | 1.5 |
| $J_D$ | $d_0$-$d_{2'}$ | 1.0 | 2.2 |
| $J_E$ | $a_1$-$a_3$, $b_0$-$b_2$ | 1.2 | 2.2 |
| $J_F$ | $a_2$-$a_3$, $d_2$-$d_3$, $d_2$-$d_{2'}$ | 1.5 | 2.2 |
| $J_G$ | $d_1$-$d_{2'}$, $d_1$-$d_3$ | 1.7 | 2.2 |
| $J_H$ | $c_0$-$c_1$ | 1.1 | 2.6 |

**Figure 11.** Carbon mass balance of the Morwell lignite sample during 300°C/9h closed pyrolysis.

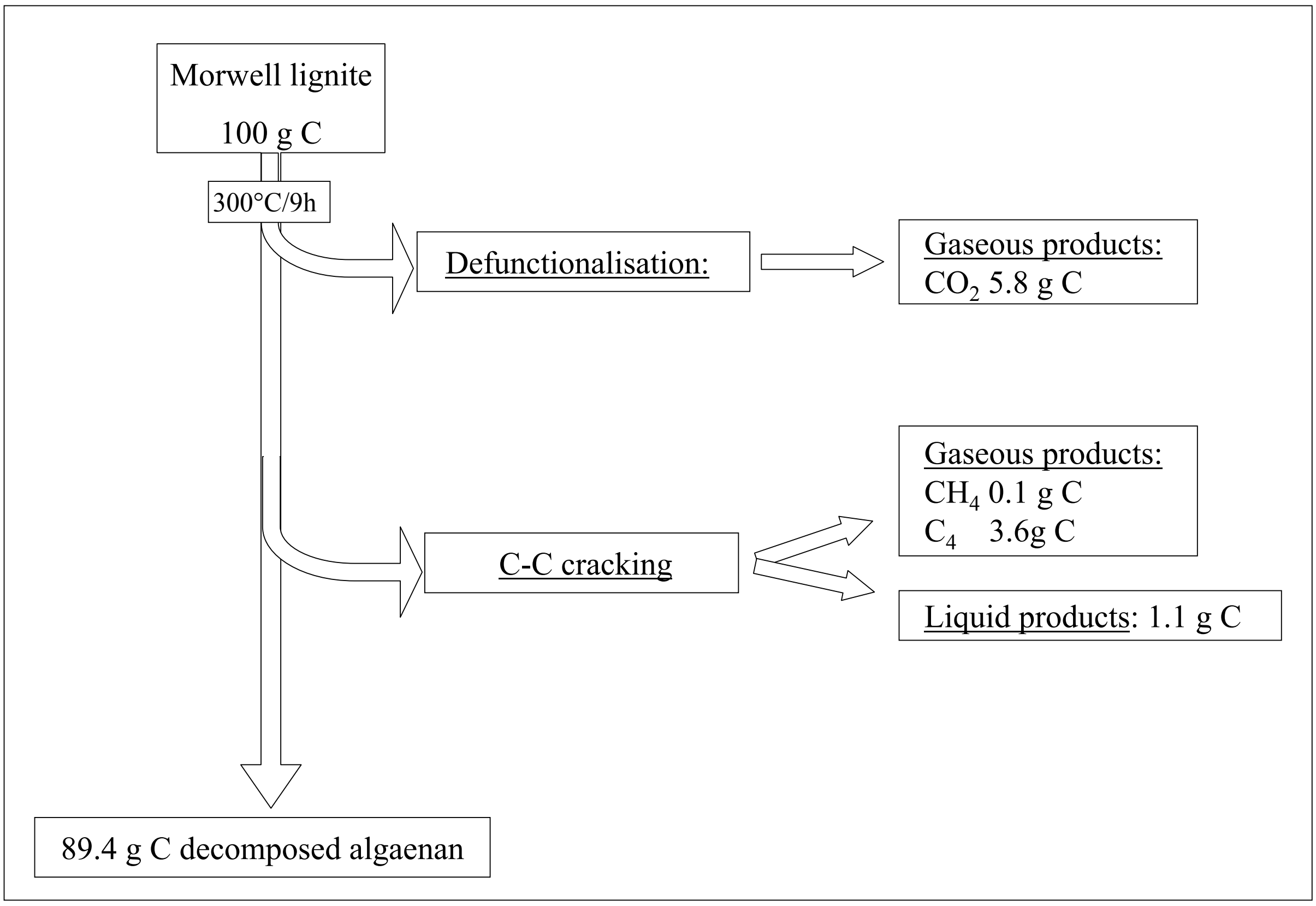